\documentclass[pra,twocolumn,superscriptaddress,showpacs,floatfix,longbibliography]{revtex4-2}
\usepackage{mathrsfs,braket}
\usepackage{amssymb, amsbsy, amsmath, latexsym, dsfont, array, layout,
graphicx,mathrsfs,color,ulem,bm}
\usepackage[colorlinks=true,citecolor=blue,urlcolor=blue]{hyperref}

\begin{document}

\title{Non-Hermitian skin effect in a spin-orbit-coupled Bose-Einstein condensate}

\author{Haowei Li}
\affiliation{CAS Key Laboratory of Quantum Information, University of Science and Technology of China, Hefei 230026, China}
\author{Xiaoling Cui}
\email{xlcui@iphy.ac.cn}
\affiliation{Beijing National Laboratory for Condensed Matter Physics, Institute of Physics, Chinese Academy of Sciences, Beijing 100190, China}
\affiliation{Songshan Lake Materials Laboratory , Dongguan, Guangdong 523808, China}
\author{Wei Yi}
\email{wyiz@ustc.edu.cn}
\affiliation{CAS Key Laboratory of Quantum Information, University of Science and Technology of China, Hefei 230026, China}
\affiliation{CAS Center For Excellence in Quantum Information and Quantum Physics, Hefei 230026, China}

\begin{abstract}
We study a Bose-Einstein condensate of ultracold atoms subject to a non-Hermitian spin-orbit coupling, where the system acquires non-Hermitian skin effect under the interplay of spin-orbit coupling and laser-induced atom loss. The presence of the non-Hermitian skin effect is confirmed through its key signatures in term of the spectral winding under the periodic boundary condition, the accumulation of eigen wavefunctions at boundaries under an open boundary condition, as well as bulk dynamics signaled by a directional flow. We show that the bulk dynamics in particular serves as a convenient signal for experimental detection. The impact of interaction and trapping potentials are also discussed based on non-Hermitian Gross-Pitaevskii equations.
Our work demonstrates that the non-Hermitian skin effect and its rich implications in topology, dynamics and beyond are well within reach of current cold-atom experiments.
\end{abstract}

\maketitle

\section{Introduction}
Non-Hermitian skin effect is an intriguing phenomena in open systems under which eigen wavefunctions exponentially localize near boundaries~\cite{WZ1}.
It has profound impact on a wide range of properties of an open system: while the non-Hermitian skin effect directly modifies the topology, spectral symmetry, and bulk dynamics in systems described by non-Hermitian effective Hamiltonians~\cite{WZ1,WZ2,murakami,ThomalePRB,Budich,mcdonald,alvarez,fangchenskin,kawabataskin,yzsgbz,stefano,tianshu,lli}, it also manifests in the long-time density-matrix dynamics driven by the master equation~\cite{wzopen,stefanoopen}.
Over the past three years, non-Hermitian skin effect and its many consequences have been experimentally confirmed in classical or photonic systems~\cite{teskin,photonskin,XDW+21,metaskin,teskin2d,scienceskin}, but not in a quantum many-body environment.

In recent studies, it is pointed out that non-Hermitian skin effect can be induced by a non-Hermitian spin-orbit coupling in cold atoms~\cite{cuisoc,yangsoc}. Therein, two hyperfine ground states of an atom are coupled by a two-photon Raman process under which a spin flip is accompanied by the change of the atomic center-of-mass momentum. Such a coupling between the atomic spin and external orbital degrees of freedom has been the subject of intense study over the past decade~\cite{SOC_1d_1,SOC_1d_2,SOC_1d_3,SOC_1d_4,SOC_2d_1,SOC_2d_3}, for its highly non-trivial influence on the system band topology~\cite{socreview6}, as well as its ability of inducing exotic few- and many-body states~\cite{socreview1,socreview2,socreview3,socreview4,socreview5}. In a very recent experiment, the Raman-induced spin-orbit coupling is further dressed by a laser-induced atom loss~\cite{joexp}. For that purpose, an additional laser selectively couples one of the hyperfine spin states to an electronically excited state, which is subsequently lost under spontaneous emission~\cite{luoexp}. For atoms that remain, their dynamics is driven by a non-Hermitian Hamiltonian featuring a non-Hermitian spin-orbit coupling and spectral singularity. While the experiment focuses on the chiral parametric transport close to the spectral singularity known as the exceptional point, theoretical studies have revealed the hidden non-Hermitian skin effect under the same configuration~\cite{cuisoc,yangsoc}. However, questions remain on amenable detection schemes, as well as the impact of interaction and trapping potentials that are present under typical experimental conditions.

In this work, we address these questions by studying a Bose-Einstein condensate of cold atoms under the experimental setup. We first confirm the results in Refs.~\cite{cuisoc,yangsoc}, demonstrating the spectral winding and the accumulation of eigen wavefunctions at boundaries by solving a single-body problem. In particular, under a periodic boundary condition (PBC), the single-body eigenspectrum of a finite-size system features closed loops on the complex plane; whereas under the open boundary condition (OBC), the eigenspectrum reduces to open arcs within the loops. Such is the topological origin of the non-Hermitian skin effect~\cite{fangchenskin,kawabataskin}. We then demonstrate a directional dynamics for wave packets in a homogeneous condensate with interactions turned off, i.e. the wavefunction propagates along the direction of the non-Hermitian spin-orbit coupling. This unidirectional propagation is the direct consequence of a persistent bulk current that is the driving force behind the namesake phenomenon of the non-Hermitian skin effect---the accumulation of wavefunctions at boundaries. Note that in our system, the bulk current, or the non-Hermitian skin effect, is due to the interplay of spin-orbit coupling and the atom loss, which can be related to a non-reciprocal inter-spin coupling via a spin rotation.

The unidirectional bulk dynamics offers a convenient signal for the detection of non-Hermitian skin effect, both for a homogeneous condensate and, more importantly, for a trapped one. We illustrate this by calculating the growth rate of the condensate~\cite{stefano,lyaexp}, which, in a homogeneous setup, corresponds to the Lyapunov exponent in the long-time limit. In particular, the peak location of the growth rate (in the so-called shift velocity) characterizes the propagation of the condensate wavefunction. Crucially, we show that in the presence of a trapping potential, a condensate initialized in the ground state at the trap center would flow along the direction of the non-Hermitian spin-orbit coupling, either in the same direction or opposite to that of the momentum transfer. The condensate wavefunction gets squeezed and eventually localized off-center, balanced by the higher potential energy there.
We further consider the impact of mean-field interactions on the unidirectional flow, and demonstrate that a repulsive (attractive) interaction enhances (suppresses) the average velocity of the flow, suggesting a stronger (weaker) non-Hermitian skin effect for a repulsively (attractively) interacting condensate. Our results illustrate bulk dynamics and the growth rate as viable signals for the experimental detection of non-Hermitian skin effect in cold atoms. Based on the flexible controls therein, it would be exciting to further explore the impact of non-Hermitian skin effect in the quantum many-body system of cold gases.

Our work is organized as follows. In Sec.~II, we present the model and characterize its single-body properties. We study the bulk dynamics of the condensate, both without and with trapping potentials in Sec.~III, by evolving the Gross-Pitaevskii equations. In Sec.~IV, we investigate the impact of interactions on the non-Hermitian skin effect. We summarize in Sec.~V.

\begin{figure}[tbp]
\includegraphics[width=9cm]{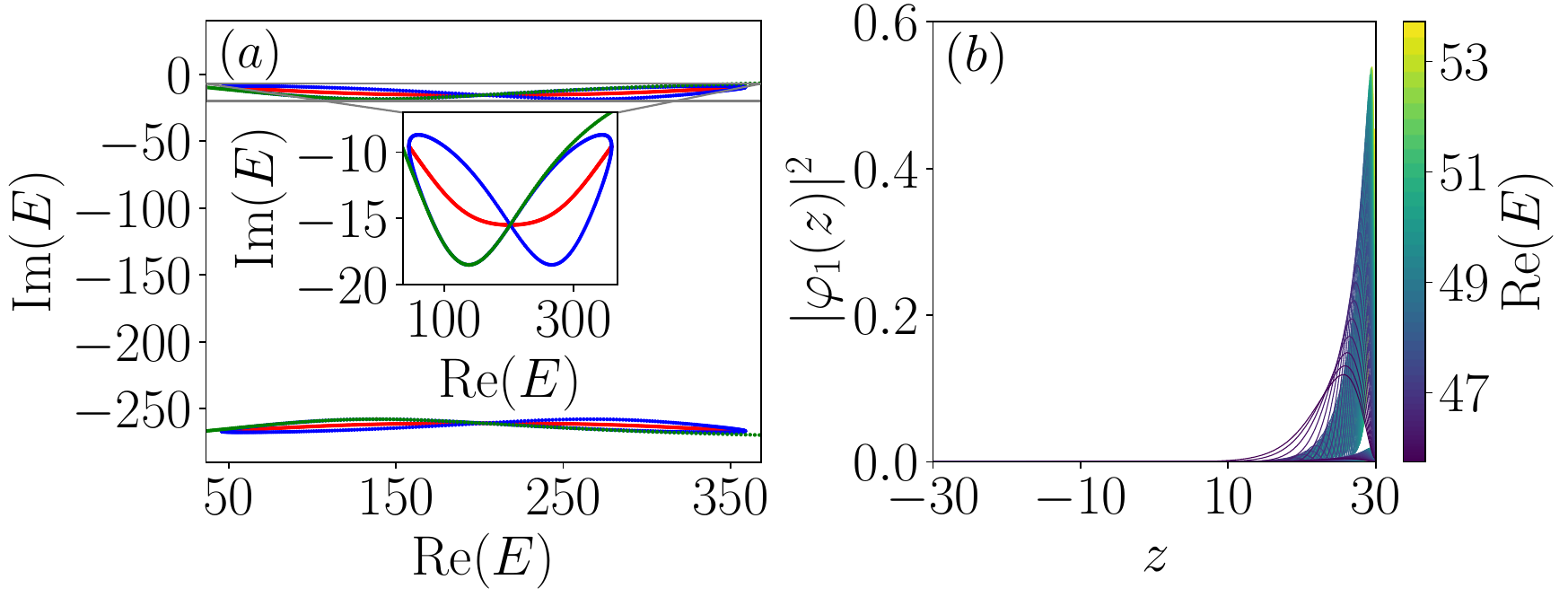}
\caption{(a) Single-particle eigenspectra of Hamiltonian (\ref{eq:Hsingle}) on the complex plane. Green: eigenspectrum of in the momentum space (an infinite system under PBC). Blue: eigenspectrum of a finite system with $z\in[-30,30]$. Red:  eigenspectrum under OBC. Inset: enlarged eigenspectra. We fix $\Omega=0.5E_r$ and $\Gamma_z=2E_r$. For calculations of finite systems, the spatial coordinates along $z$ are discretized into $480$ segments.
(b) Spatial distribution of the $100$ eigenstates with the smallest real components (indicated by the color bar).
}
\label{fig:fig1}
\end{figure}

\begin{figure}[tbp]
\includegraphics[width=9cm]{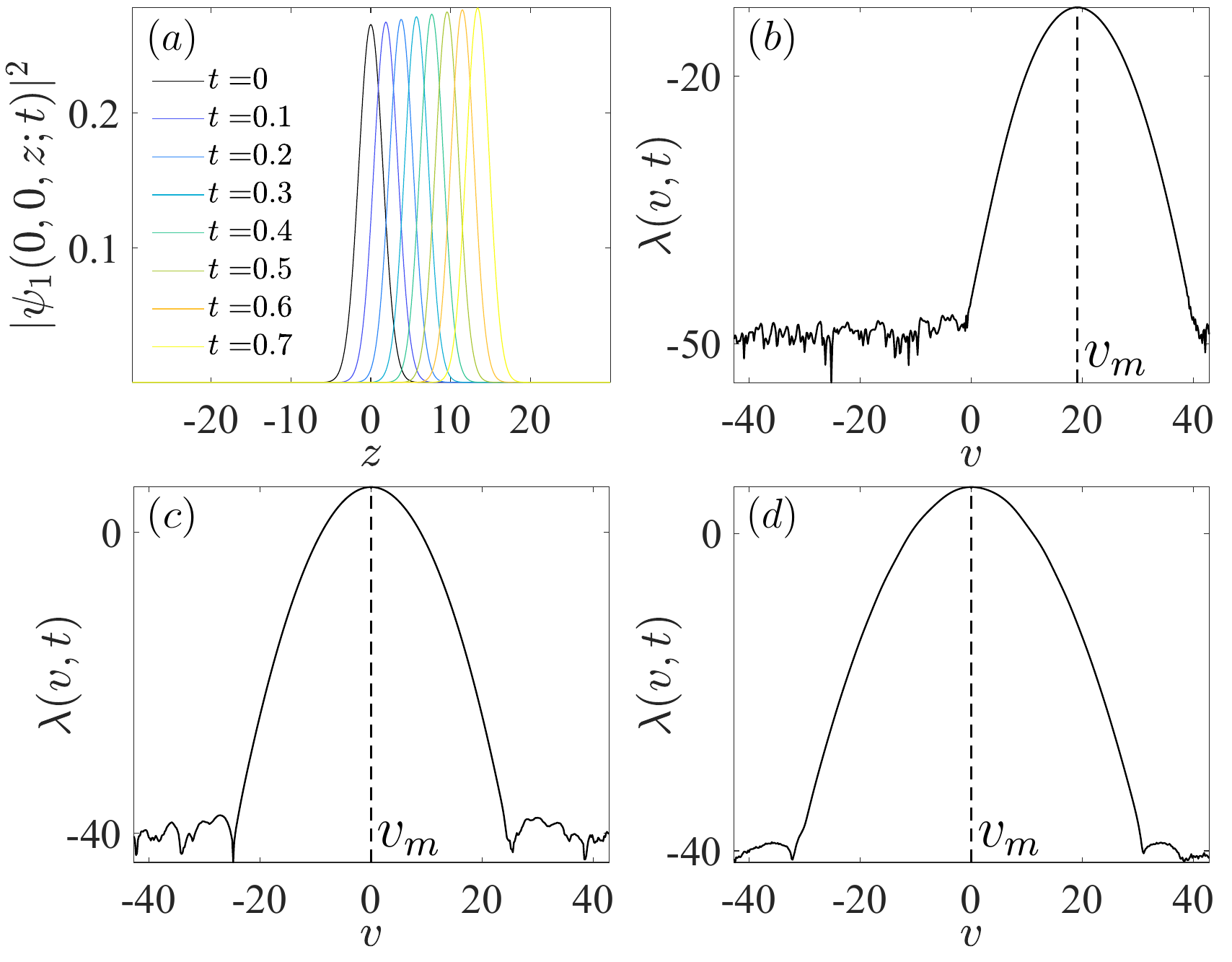}
\caption{(a) Propagation of the condensate wavefunction in the bulk, with $\Omega=0.5E_r$ and $\Gamma_z=2E_r$.
(b) Growth rate as a function of the shift velocity under the parameters of (a). (c) Growth rate with $\Omega=0$ and $\Gamma_z=2E_r$. (d) Growth rate with $\Omega=0.5E_r$ and $\Gamma_z=0$.
}
\label{fig:fig2}
\end{figure}

\begin{figure*}[tbp]
\includegraphics[width=15cm]{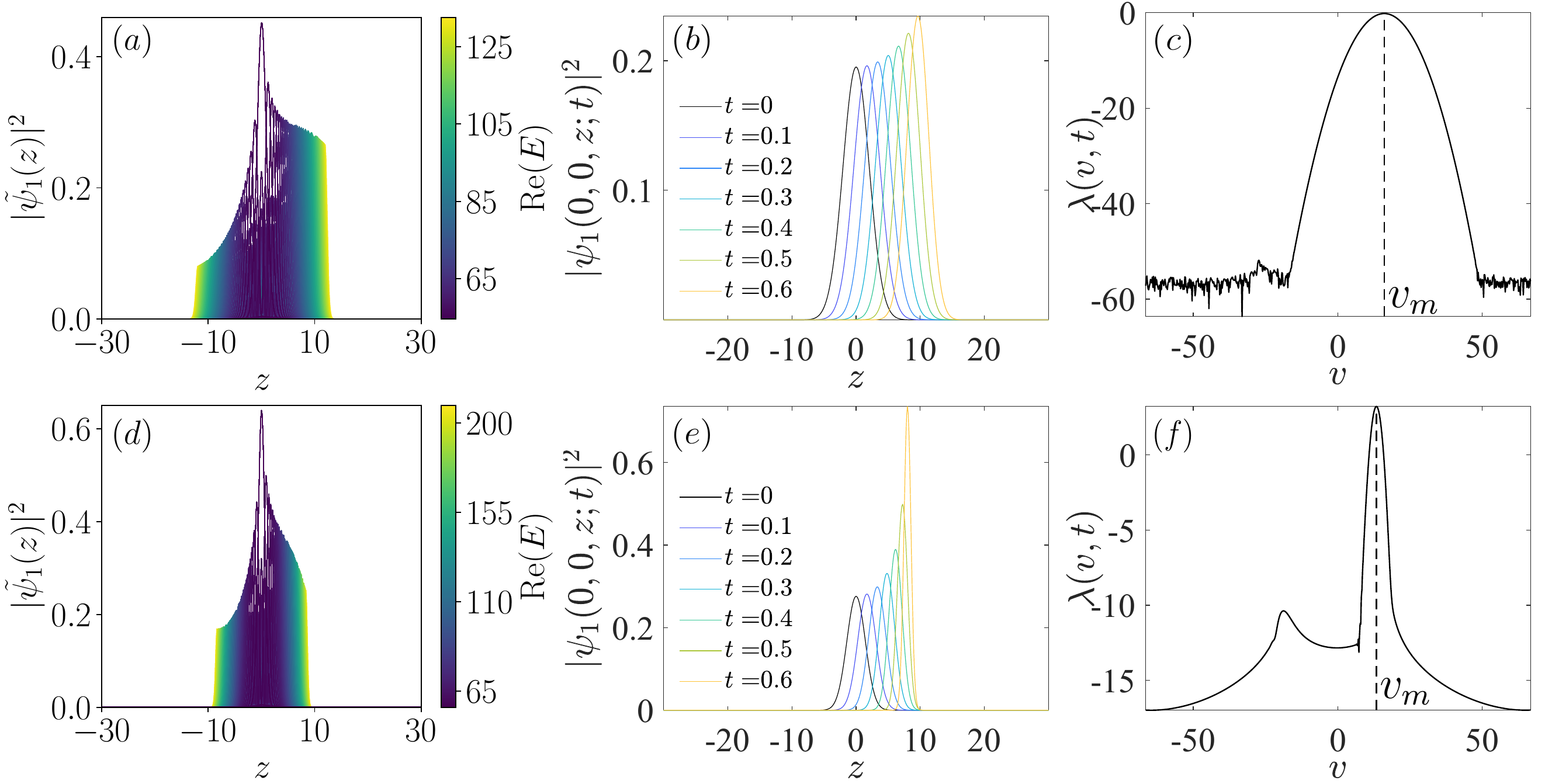}
\caption{(a)(d) Spatial distribution of eigen wavefunction along the $z$ direction in a homogeneous harmonic trap, with $\Omega=0.5E_r$ and $\Gamma_z=5E_r$.
For the numerical calculations here, we take a cylindrical coordinate, discretizing $z\in[-30,30]$ into $480$ segments, and the radial coordinate $\rho\in[0,4]$ into $8$ segments. We plot the radial-integrated spatial distribution of the $800$ eigenstates with the smallest real components, colored according to $\text{Re}(E)$ (see color bar).
 Specifically, $\tilde{\psi}_1(z)=2\pi \int \rho d\rho \psi_1(\rho,z)$.
(b)(e) Propagation of the condensate wavefunction in the bulk.
(c)(f) Growth rate as a function of the shift velocity. The peak shift velocity $v_m\approx 16.04$ in (c) and $v_m\approx 13.33$ in (f).
The trapping potential is $\omega=\omega_0$ in (a)(b)(c), and $\omega=2\omega_0$ in (d)(e)(f).
}
\label{fig:fig3}
\end{figure*}

\section{Model}

We consider the recently implemented non-Hermitian spin-orbit coupling in cold atoms~\cite{joexp}, where the single-body Hamiltonian is given by
\begin{align}
H&=H_0+\frac{i}{2}(\sigma_z-1)\Gamma_z\nonumber\\
&=\frac{1}{2m}\left(-i\hbar\nabla +\hbar k_r \bm{e}_z\sigma_z \right)^2+\Omega \sigma_x+\frac{i}{2}(\sigma_z-1)\Gamma_z.
\label{eq:Hsingle}
\end{align}
Here $\Omega$ and $2\hbar k_r$ are respectively the effective Rabi frequency and momentum transfer of the Raman process, $\sigma_{x,y,z}$ are the Pauli operators for the two hyperfine spin species, $m$ is the atomic mass, and $\Gamma_z$ is the laser-induced loss rate for the spin-down atoms.

The non-Hermitian Hamiltonian derives from the conditional dynamics of the Lindblad equation $d\rho/dt=-\frac{i}{\hbar} H_0\rho+\frac{i}{\hbar} \rho H_0^\dag+\Gamma_z S\rho S^\dag$, where $\rho$ is the atomic density matrix and $S$ is the quantum jump operator describing the laser-induced, spin-selective loss. Specifically, the dynamics of the system is driven by the non-Hermitian Hamiltonian $H$, when the quantum jump terms are dropped~\cite{QJ}. Experimentally, such a conditional dynamics is implemented by probing only atoms that remain in the system. The description is exact in the absence of interactions, and when the spontaneous emission back into the two hyperfine spin states can be neglected.

For our numerical simulations below, we use parameters that are of similar magnitude compared to those in spin-orbit coupled $^{87}$Rb atoms~\cite{SOC_1d_2}. For instance, we take $E_0=\hbar\omega_0$ as the unit of energy, with $\omega_0=100$Hz. Correspondingly, the unit of time is $1/\omega_0\approx 10$ms, and the unit of length $x_0=\sqrt{\hbar/m\omega_0}$. For the spin-orbit-coupling parameters, we take the recoil energy $E_r\approx 2\pi \times 2.2$kHz, which corresponds to $E_r/E_0=44\pi$ and $k_rx_0=16.62$.
%We further fix $\Omega=0.5E_r$ and $\Gamma_z=E_r$ throughout the work.

While the experiment focuses on the parity-time symmetry and spectral singularity (the exceptional point) of the Hamiltonian (\ref{eq:Hsingle}), the system also has non-Hermitian skin effect, despite being a continuous model. To confirm this point, we calculate the single-particle eigenspectra of Eq.~(\ref{eq:Hsingle}) under both PBC and OBC, for a one-dimensional gas along the $z$ direction---the direction of the spin-orbit coupling.

For a one-dimension gas, the single-particle eigenspectrum in the momentum space is given by
\begin{align}
E_\pm(k)=&E_k+E_r-i\frac{\Gamma_z}{2}\nonumber\\
&\pm\sqrt{4E_kE_r+2i\sqrt{E_kE_r}\Gamma_z-\frac{\Gamma_z^2}{4}+\Omega^2},
\end{align}
where $E_k=\hbar^2 k^2/2m$, $k$ is the momentum along the $z$ direction.
The eigenspectrum corresponds to the green curves in Fig.~\ref{fig:fig1}, where the two branches $E_\pm(k)$ are self-connected at infinite $|\text{Re}E|$, forming closed loops on the complex energy plane. For a finite system but still under PBC, the loop structures are easier to see (blue curves).
This is consistent with the spectral winding, the topological origin of the non-Hermitian skin effect. By contrast, under an OBC, the eigenspectra collapse into open arcs (red curves) within the spectral loop under the PBC. Correspondingly, the spatial distribution of the eigen wavefunctions accumulate at the open boundary, as shown in Fig.~\ref{fig:fig1}(b). Here the position of the localization (the left or right boundary) is tunable through parameters of the spin-selective loss or the spin-orbit coupling.
Note that we denote the wavefunctions for the two spin species as $\varphi_{1,2}(z)$, which are normalized according to $\int [|\varphi_1(z)|^2+|\varphi_2(z)|^2] dz=1$.

\section{Dynamic signal of the non-Hermitian skin effect}

In cold atomic gases, a sharp open boundary is typically difficult to engineer, and eigenspectrum is not easy to probe, unlike classical or photonic simulators. Therefore, the most ostensible signatures of the non-Hermitian skin effect, as demonstrated in Fig.~\ref{fig:fig1}, can be experimentally elusive. However, systems with non-Hermitian skin effect also possess unique signatures in the bulk dynamics, which, as we reveal in this section, serve as convenient dynamic signals of the non-Hermitian skin effect.

We first consider the non-interacting case, where the post-selection principle underlying the non-Hermiticity of the Hamiltonian is valid, and atoms that remain in the system evolve according to (\ref{eq:Hsingle}). In Fig.~\ref{fig:fig2}(a), we show the numerically simulated propagation of the condensate wavefunction. The initial state is a Guassian wave packet $\psi_{1,2}(x,y,z;t=0)\propto \exp[-(x^2+y^2+z^2)/w^2]$, with $w=3$ and normalized to unity. We evolve the wavefunction in real space using the Schr\"odinger's equation
\begin{align}
i\hbar \frac{d}{dt}\left(\begin{array}{c}
\psi_{1}\\
\psi_{2}
\end{array}\right)=H\left(\begin{array}{c}
\psi_{1}\\
\psi_{2}
\end{array}\right),
\end{align}
where $\psi_{1,2}$ are the wavefunctions for the two spin species. For numerical calculations, we discrete the spatial coordinate along $z$ into $480$ segments in the range $z\in[-30,30]$, and those along $x$ and $y$ each into $32$ segments in the range $x,y\in [-12,12]$.
The spatial derivatives in the Hamiltonian are then translated into finite differences.

In Fig.~\ref{fig:fig2}(a), we show the evolution of the wavefunction along the $z$ axis, with $(x=0,y=0)$.
The directional propagation of the wavefunction suggests a persistent bulk current that lies at the origin of the non-Hermitian skin effect. In previously studied lattice models, the non-Hermitian skin effect is often associated with a non-reciprocal hopping~\cite{WZ1}.
% in the form of $\Omega\sigma_x+i\Gamma\sigma_y$ when translated into a pseudospin basis.
It is important to note that the non-Hermitian spin-orbit coupling with a manifestly spin-selective loss here $\Omega\sigma_x+i\frac{\Gamma_z}{2}\sigma_z$, is related to a non-reciprocal model through a spin rotation $U=e^{i\frac{\pi}{4}\sigma_x}$.

To provide a quantitative measure of the unidirectional propagation, we define a wavefunction growth rate
\begin{align}
\lambda(v,t)=\frac{\ln |\psi_1(0,0,z=vt;t)|}{t},
\end{align}
where $v$ is the shift velocity. In the long-time limit $t\rightarrow\infty$, the growth rate converges to the Lyapunov exponent of a homogeneous system~\cite{stefano}.
As shown in Fig.~\ref{fig:fig2}(b), the peak of $\lambda(v,t)$ lies at a finite shift velocity $v_m\approx 19.1$, which is essentially the propagation velocity of the wave-packet peak in Fig.~\ref{fig:fig2}(a). By contrast, with a vanishing spin-orbit coupling ($\Omega=0$), or a vanishing atomi loss ($\Gamma_z=0$), the growth rate peaks at $v_m=0$ [see Fig.~\ref{fig:fig2}(c)(d)], indicating the absence of the bulk current and the non-Hermitian skin effect. Note that without a trapping potential, the peak location of $\lambda(v,t)$ already converges to a finite $v_m$ for our numerical simulations of a finite time evolution.

Experimentally, condensates are typically subject to a harmonic trapping potential, which provides a natural boundary condition, though not as sharp as an ideal OBC.
For a condensate initialized at the ground state, i.e., near the center of the trap, we expect the directional flow to persist in systems with non-Hermitian skin effect. However, when the condensate moves off-center, the soft boundary that is the harmonic trap would impact the wavefunction propagation and eventually stop it near the edge of the trap.

Such an intuitive picture is indeed confirmed in Fig.~\ref{fig:fig3}, where the dynamics is governed by
\begin{align}
i\hbar \frac{d}{dt}\left(\begin{array}{c}
\psi_{1}\\
\psi_{2}
\end{array}\right)=[H+V(r)]\left(\begin{array}{c}
\psi_{1}\\
\psi_{2}
\end{array}\right),
\end{align}
where $V(r)=\frac{1}{2}m\omega^2 r^2$ with the trapping frequency $\omega$.

First, in Fig.~\ref{fig:fig3}(a)(d), we show typical spatial distribution (along the $z$ axis) of eigenstate wavefunctions for $H+V(r)$, with trapping frequencies $\omega=\omega_0$ and $\omega=2\omega_0$, respectively. The off-center distribution is a direct manifestation of the non-Hermitian skin effect in a harmonic trap.

We then numerically evolve the ground state, where a directional propagation of the wavefunction is observed [see Fig.~\ref{fig:fig3}(b)(e)], consistent with dynamics in the homogeneous case of Fig.~\ref{fig:fig2}. However, in a trapping potential, the directional propagation would slow down and eventually be stopped by the trap edge, which is apparent by comparing dynamics under different trapping frequencies Fig.~\ref{fig:fig3}(b)(e). What we observe here is essentially the dynamic accumulation of wavefunctions at boundaries, driven by the non-Hermitian skin effect.

In Fig.~\ref{fig:fig3}(c)(f), we show the growth rates under different trapping frequencies. At short times, the growth rate is peaked at a finite $v_m$, consistent with the presence of non-Hermitian skin effect.
After a sufficiently long time, another peak emerges in $\lambda(v,t)$ [see Fig.~\ref{fig:fig3}(f)], indicating the backflow of condensate atoms as they are reflected from the boundary (the boundary to the right in this case).
While such an effect is more apparent for deeper traps [compare Fig.~\ref{fig:fig3}(c)(f)], in the long-time limit, the peak velocity $v_m$ of the growth rate uniformly drops to zero, as the propagation is stopped by the boundary.

As such, a trapped condensate offers an intriguing scenario for the dynamic detection the non-Hermitian skin effect, where the short-time dynamics is dominated by the directional bulk current, while the long-time dynamics is dominated by the accumulation of atoms toward the boundary.

\section{Interaction effect}

\begin{figure}[tbp]
\includegraphics[width=9cm]{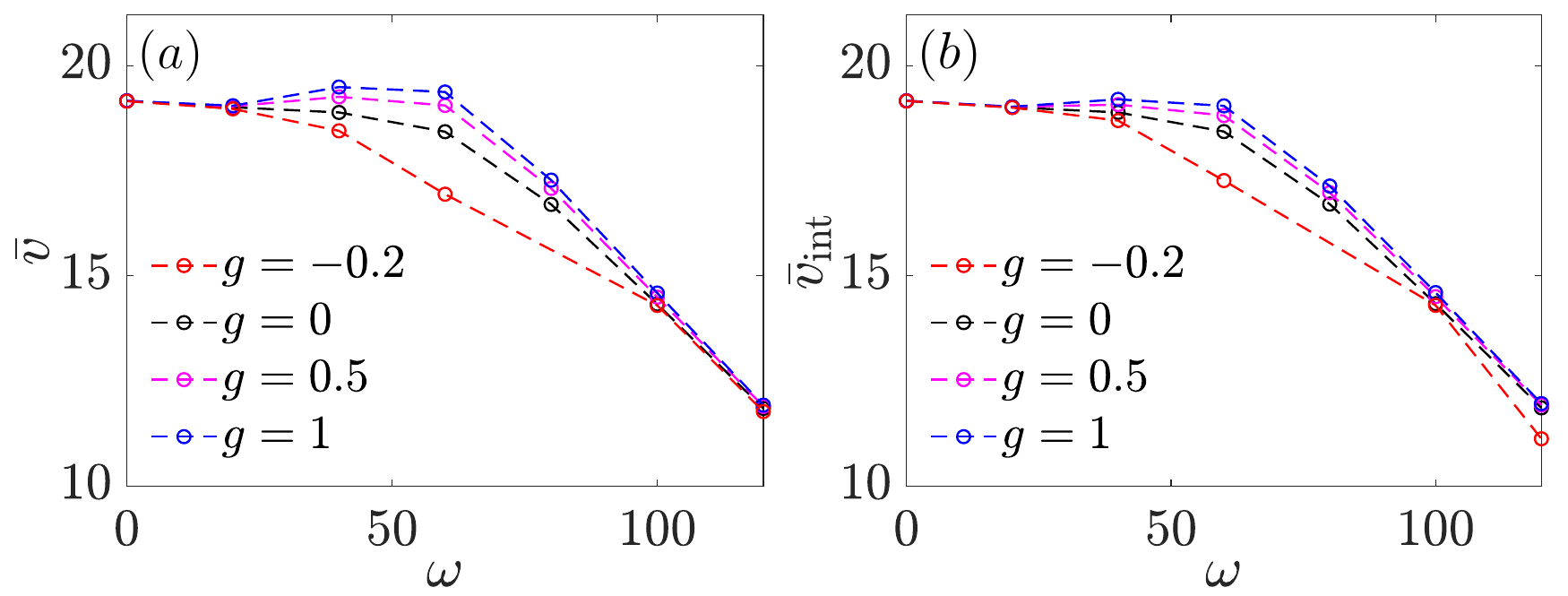}
\caption{Effect of condensate interaction on the non-Hermitian skin effect in a trapped gas. See main text for the definition of the average propagation speed $\bar{v}$ in (a), and the integrated propagation speed $\bar{v}_{\text{int}}$ in (b). Other parameters are the same as those in Fig.~\ref{fig:fig2}(a)(b).
}
\label{fig:fig4}
\end{figure}

We now consider the impact of interaction, particularly on the dynamic signal of the wavefunction propagation. While the non-Hermitian description based on post-selection can be questionable when interatomic interactions are considered, we assume that the condensate is always in a coherent state, which is an eigenstate of the jump operator and hence a stead state of the dynamics. We may then write down the Gross-Pitaevskii equation
\begin{align}
i\hbar \frac{d}{dt}\left(\begin{array}{c}
\psi_{1}\\
\psi_{2}
\end{array}\right)=[H+V(r)+H_{\text{int}}]\left(\begin{array}{c}
\psi_{1}\\
\psi_{2}
\end{array}\right),
\end{align}
with the mean-field interaction term given by
\begin{align}
H_{\text{int}}=\left(\begin{array}{cc}
g|\psi_1|^2&0\\
0&g|\psi_2|^2
\end{array}\right).
\end{align}
Here $g$ is the interaction rate, and we have neglected inter-species interactions.
For numerical simulations, the unit of interaction rate is taken as $g_0=4\pi\hbar^2 a_s/m$, where $a_s=100 a_0$ ($a_0$ is the Bohr radius).
We initialize a condensate of $N=5\times 10^5$ atoms, in the absence of atom loss and spin-orbit coupling. This is achieved using the imaginary time evolution.
During the time evolution, the atomic number continuously decreases due to loss. The interaction effect therefore also becomes diminishingly small with the passage of time.

In Fig.~\ref{fig:fig4}(a), we show the numerically simulated average velocity of the wavefunction propagation along the $z$ axis with $(x=0,y=0)$, which is defined as
\begin{align}
\bar{v}=\frac{\int  z|\psi_1(0,0,z;t)|^2 dz}{t\int  |\psi_1(0,0,z;t)|^2 dz},
\end{align}
where we fix $\omega_0 t=1.2$ for the time evolution.
When the trapping frequency vanishes, the average velocity converges, regardless of the interaction strength $g$. This is consistent with the convergence of $v_m$ in Fig.~\ref{fig:fig2}(a)(b). By contrast, for sufficiently large trapping frequency (or sufficiently long evolution time), $\bar{v}$ vanishes, indicating the accumulation of wavefunction at the trap edge, consistent with the results in Fig.~\ref{fig:fig3}.
But in general, a repulsive (attractive) interaction would facilitate (suppress) the wavefunction propagation, as indicated by Fig.~\ref{fig:fig4}(a).
This is a direct evidence of the interplay of interaction and non-Hermitian skin effect.

While $\bar{v}$ is directly related to the growth rate in Fig.~\ref{fig:fig3}, it is typically difficult to measure experimentally. Instead, in Fig.~\ref{fig:fig4}(b), we calculate the integrated propagation speed $\bar{v}_{\text{int}}$, defined as
\begin{align}
\bar{v}_{\text{int}}=\frac{\int  z|\psi_1(x,y,z;t)|^2 dx dy dz}{t\int  |\psi_1(x,y,z;t)|^2 dx dy dz}.
\end{align}
The results are qualitatively consistent with those in Fig.~\ref{fig:fig4}(b), whereas $\bar{v}_{\text{int}}$ can be experimentally detected by measuring the overall propagation of the condensate along the direction of the spin-orbit coupling.

\section{Conclusion}

We show that non-Hermitian skin effect emerges in a condensate of cold atoms under synthetic spin-orbit coupling and laser-induced atom loss. The non-Hermitian skin effect can be dynamically detected through the directional propagation of wavefunctions in the bulk, as well as the dynamic accumulation of atoms near the edge of the harmonic trapping potential. We also demonstrate that the mean-field interaction can have a detectable impact on the directional propagation, offering an experimentally relevant example wherein the interplay of interaction and non-Hermiticity can be directly probed. For future studies, it is interesting to examine the manifestation of non-Hermitian skin effect in a Fermi gas, where the effect of Fermi-Dirac statistics on the non-Hermitian skin effect can be systematically studied.

\section*{Acknowledgements}

This work has been supported by the National Key Research and Development Program of China (2018YFA0307600, 2017YFA0304100), the National Natural Science Foundation of China (No. 11974331,No.12074419), and the Strategic Priority Research Program of Chinese Academy of Sciences (No. XDB33000000).

\bibliographystyle{apsrev4-1}

\end{document}